\documentclass[a4paper]{article}

\newcommand{\ctg}{\ensuremath{\text{ctg}}}

\usepackage{amsmath}

\title{Canonical quantization and black hole perturbations\thanks{Talk presented by Zolt\'an Perj\'es}}
\makeatletter
\author{\framebox{Zolt\'an Perj\'es} and \'Arp\'ad Luk\'acs\\
    KFKI Research Institute for Particle and Nuclear Physics,\\
    Budapest 114, P.O. Box 49, H-1525 Hungary\\
    arpi@rmki.kfki.hu}

\begin{document}
\maketitle

\begin{abstract}
We examine the possibility of a constraint-free quantization of linearized gravity, based
on the Teukolsky equation for black hole perturbations.

We exhibit a simple quadratic (but complex) Lagrangian for the Teukolsky equation, leading to the interpretation that the elementary
excitations (gravitons bound to the Kerr black hole) are unstable. 
\end{abstract}
\begin{flushleft}
Classification: 04.60.Ds, 04.70.-s, 04.25.Nx
\end{flushleft}

\section{Introduction}
Quite a number of attempts have been made and are being made to quantize Einstein's gravitation theory. One of the earliest attempts to
use canonical quantization of linearized gravity goes back to Gupta \cite{Gupta1}. The canonical approach to quantum gravity using the 3+1
decomposition has been worked out by Arnowitt, Deser and Misner \cite{ADM}. A different idea has been put forward by Penrose
\cite{Penrose} introducing twistors.
Ashtekar \cite{Ashtekar} introduced spinor variables to handle the constraints. Nowadays loop quantum gravity is being actively pursued,
see the review by Rovelli \cite{Rovelli}.  In spite of the remarkable progress achieved using
these beautiful constructions, there remains a lot to be done.

Undoubtedly most of the difficulties in all canonical approaches to quantum gravity come from the constraints.
Another problem, closely related to the treatment of constraints is the choice of the canonical coordinates. A good choice of
coordinates leads to less constraint equations, and thus it makes quantization easier.

In general relativity, the physical state of space-time is described by a symmetric, second rank tensor that can be parametrized by ten
components. Due to the constraints, we should introduce less 
than ten independent canonical coordinates, and their conjugate momenta.

The well known Teukolsky equation \cite{Teuk1}, which is a linear PDE, describes the perturbations of a Kerr space-time
with a single scalar quantity, $\psi$, one of the Newman--Penrose spin
coefficients.

Chrzanowski, Misner \cite{Chrz} and Ori \cite{Ori} have shown the way to reconstruct the metric perturbations from the solution of
the Teukolsky equation by acting on it with a differential operator.

This makes the quantity $\psi$ a promising candidate as a canonical coordinate in a linearized quantum
gravity, provided a Hamiltonian formulation of the Teukolsky equation
(\ref{eq:teuk}) can be given. This problem will be examined in the next section.

One would think, that the easiest way to obtain a canonical formalism based on the Teukolsky equation is to reconstruct the perturbations of
the metric tensor, and subsequently the quadratic part of the corresponding perturbed Lagrangian. This could then be taken
as a starting point for a canonical quantization. However,
the formulae for the metric perturbations are very complicated, and contain fourth order derivatives of $\psi$.

One could just as well try to find a Lagrangian directly for Teukolsky's equation in itself, and use that for quantization. Unfortunately,
as we will see later, there is no quadratic, first order real function, which gives (\ref{eq:teuk}) as an Euler--Lagrange equation.

We have identified a complex Lagrangian for the Teukolsky equation. A possible solution to quantize the Teukolsky equation could be
considering the complex part of the Lagrangian as an interaction
term. The interaction-free part can be then simply quantized, and it's quanta may be identified as gravitons. As a result of the
perturbations (the complex part of the Lagrangian), these excitations decay.

\section{Is a Lagrangian formulation of  Teukolsky's equation possible?}
Some basic features of the Teukolsky equation can be found in the Appendix. Here we recall that this equation governs the dynamics of one
of the Riemann tensor's tetrad components in a linearized approximation over 
a Kerr background.

Using the notations of the Appendix, Teukolsky's equation \cite{Teuk1} can be written as
\begin{equation}\label{eq:teuk}\begin{aligned}
\left[\frac{(r^2+a^2)^2}{\Delta}-a^2\sin^2 \vartheta \right]&\frac{\partial^2\psi}{\partial
                                                               t^2}+\frac{4Mar}{\Delta}\frac{\partial^2\psi}{\partial t\partial\varphi}
                                                               +\left[\frac{a^2}{\Delta}-\frac{1}{\sin^2 \vartheta
                                                               }\right]\frac{\partial^2\psi}{\partial\varphi^2}\\
-\Delta^{-s}\frac{\partial}{\partial r}\left(\Delta^{s+1}\frac{\partial\psi}{\partial r}\right)&-\frac{1}{\sin \vartheta
                                                               }\frac{\partial}{\partial\vartheta}\left(\sin \vartheta \frac{\partial
                                                               \psi}{\partial
    \vartheta}\right) -2s\left[\frac{a(r-M)}{\Delta}+\frac{i\cos \vartheta }{\sin^2 \vartheta }\right]\frac{\partial\psi}{\partial\varphi}\\
-2s&\left[\frac{M(r^2-a^2)}{\Delta}-r-ia\cos \vartheta \right]\frac{\partial\psi}{\partial t}+(s^2\ctg^2 \vartheta-s)\psi=4\pi\Sigma T,
\end{aligned}\end{equation}
where $T$ is a source term, calculated from the energy--momentum  tensor of other fields. In the following, we concentrate on the free gravity case ($T=0$).

Over a Kerr geometry, this equation describes massless scalar fields ($s=0$), electromagnetic perturbations ($s=\pm 1$) and gravity ($s=\pm 2$).

In this section, we examine, whether it is possible to obtain equation (\ref{eq:teuk}) as an Euler--Lagrange equation from a {\sl real} Lagrangian of the form
\begin{equation}\label{eq:lagfgv}
{\mathcal L}=ag^{ik}\overline{\partial_i (b\psi)}{\partial_k (b\psi)}+c\overline{\psi}\psi,
\end{equation}
which is the most general first order quadratic local Lagrangian of a scalar field. The motivation to try to find such a Lagrangian (eg.\
local, not containing higher derivatives) is the possibility to quantize it easily.

In order for ${\mathcal L}$ be real, $a$ and $c$ have to be real, and $b$ complex arbitrary functions of the coordinates. From the above
function, the Euler--Lagrangian equation is given by the formula
\begin{equation}\label{eq:eulag}
\partial_i\frac{\partial \sqrt{-q}{\mathcal L}}{\partial\partial_i\overline{\psi}}-\frac{\partial{\sqrt{-g}\mathcal L}}{\partial{\overline\psi}} =0,
\end{equation}
where
\[\sqrt{-g}=(r^2+a^2\cos^2 \vartheta )\sin \vartheta ,\]
is the determinant of the metric tensor.

Calculating the above equation, and taking the coefficients of $\phi$'s derivatives in (\ref{eq:eulag}) and (\ref{eq:teuk}) yields
equations for $a$, $b$, $c$ and their derivatives. When trying to solve
these equations, one finds, that the integrability conditions (eg.\ the $r$ derivative of $\partial a/\partial t$ should be equal to the
$t$ derivative of $\partial a/\partial r$) cannot be fulfilled. This clearly shows that no real Lagrangian of the form (\ref{eq:lagfgv})
can lead to Teukolsky's equation.

While this work was in progress, we have learnt from the summary of Stephen Anco's talk at the Montreal Workshop on the Interaction of Gravity
with External Fields \cite{Anco1}, that
Teukolsky's equation has no Lagrange function. According to
Anco \cite{Anco1}, this is because the non self-adjointness of (\ref{eq:teuk}), which means that
\[\int \overline{\phi} {\sf D}\psi \ne \overline{\int \overline{\psi}{\sf D}\phi},\]
where $D$ denotes the differential operator in (\ref{eq:teuk}).

To show that the equation is indeed not self-adjoint, one should take the terms in $\int \overline{\phi} {\sf D}\psi$ and do the partial integrations to calculate its adjoint, and compare it with the
complex conjugate of $\int \overline{\psi} {\sf D}\phi$.

In the first term, self-adjointness is obvious: the coefficient of $\partial^2/\partial t^2$ does not depend on $t$. Similar is the case with the second and the third term.

However, the non-self-adjointness can be seen on the fourth term:
\begin{equation}\label{eq:negytag}
\int \overline{\phi}\Delta^{-s}\frac{\partial}{\partial r}\left(\Delta^{s+1}\frac{\partial\psi}{\partial r}\right),
\end{equation}
where substituting $\Delta$, and expanding the derivatives yields in the $s=2$ case
\[\int\phi\frac{3(r^2-2Mr+a^2)^2(2r-2M)\frac{\partial\psi}{\partial r}+(r^2-2Mr+a^2)^3\frac{\partial^2\psi}{\partial r^2}}{(r^2-2Mr+a^2)^2},\]
which takes the form
\begin{equation}
\int\psi\left[(r^2-2Mr+a^2)\frac{\partial^2\phi}{\partial r^2}-(2r-2M)\frac{\partial \phi}{\partial r}-4\phi\right]
\end{equation}
when doing the partial integrations. One can see, that exchanging the roles of $\psi$ and $\phi$ in the original integral, and conjugating
does not give such terms, and no other term can cancel the $r$-derivatives in the above.

Bini, Cherubini and Jantzen \cite{Bini} gave another form of the above equation, closely resembling an ordinary Klein--Gordon equation. They
have proven, that the Teukolsky equation is a generalization of the Klein--Gordon equation for arbitrary spin weight $s$. Examining the
following form of the equation, it is easy to point out the cause of the non self-adjointness:
\begin{equation}\label{eq:Bini}
\left[(\nabla^i+s\Gamma^i)(\nabla_i+s\Gamma_i)-4s^2\Psi_2^A\right]\psi=4\pi T,
\end{equation}
where $\Gamma$ is a four-vector with components
\begin{equation}\label{eq:gamma}\begin{aligned}
\Gamma^{\,t} &= -\frac{1}{\Sigma}\left[\frac{M(r^2-a^2)}{\Delta}-(r+ia\cos \vartheta )\right]\\
\Gamma^{\,r} &= -\frac{1}{\Sigma}(r-M)\\
\Gamma^{\,\vartheta} &= 0\\
\Gamma^{\,\varphi} &= -\frac{1}{\Sigma}\left[\frac{a(r-M)}{\Delta}+i\frac{\cos \vartheta }{\sin^2 \vartheta }\right].
\end{aligned}\end{equation}
It is easy to show that
\[\nabla_i \Gamma^i = -\frac{1}{\Sigma}\quad\text{és}\quad \Gamma_i \Gamma^i = \frac{1}{\Sigma}\ctg^2 \vartheta+4\Psi_2^A.\]

Here $\Psi_2^A$ denotes the value of the  $\Psi_2$ spin coefficient in an unperturbed Kerr space-time,
\[\Psi_2^A=-\frac{M}{(r-ia\cos \vartheta )^3}.\]

Now we can see, that the equation is non self-adjoint, because the $\Gamma_i$-components are not purely imaginary, unlike the Klein--Gordon
case, where $\Gamma_i=iA_i$, where $A_i$ is the electromagnetic four-potential. These complex terms make the equation absorptive.

\section{Perturbative quantization of the  Teukolsky equation}\label{se:pkv}
In the previous section we have seen that a direct canonical quantization of the Teukolsky equation is not possible because of the
absorptive terms in the Lagrangian. On the other hand, the  Bini--form (\ref{eq:Bini}) of the Teukolsky equation (\ref{eq:teuk}) is in close
resemblance to the Klein--Gordon equation. This suggests the following solution: we separate the absorptive part of the equation to treat it
as a perturbation, and this way we can construct a Lagrangian for the self-adjoint part. This will be real, and so the Fock space
construction is well defined, a particle interpretation can be given.

Introducing the notation of
\begin{equation}\label{eq:ab}
\begin{aligned}
A_i &= \Im (\Gamma_i)\\
B_i &= \Re (\Gamma_i),
\end{aligned}
\end{equation}
the Bini form of the equation can be written as
\begin{equation}\label{eq:biniab}
\left[(\nabla^i+s(B^i+iA^i))(\nabla_i+s(B_i+iA_i))-4s^2\Psi_2^A\right]\psi=4\pi T,
\end{equation}
and if we keep only the $A_i$s, we get an equation which is formally a Klein--Gordon equation in an external electromagnetic field. The
(complex) Lagrangian for this equation is:
\begin{equation}\label{eq:teuklag}
\mathcal{L}=\left\{\nabla^i-s(B^i+iA^i)\right\}\overline{\psi}\left\{\nabla_i+s(B_i+iA_i)\right\}\psi+4s^2\Psi_2^A\overline{\psi}\psi,
\end{equation}
which can be written as a sum of an unperturbed part, and a perturbation term:
\[\mathcal{L}=\mathcal{L}_{KG}+\mathcal{L}_1,\]
where
\begin{equation}\label{eq:kglag}
\mathcal{L}_{KG}=\overline{(\nabla^i+isA^i)\psi}(\nabla_i+isA_i)\psi
\end{equation}
is the Klein-Gordon Lagrangian, and
\begin{equation}\label{eq:plag}
\mathcal{L}_1=-s(B^i\overline{\psi}\nabla_i\psi-B^i\psi\nabla_i\overline{\psi})+2isB^iA_i\overline{\psi}\psi+4is^2\Psi_2^A\overline{\psi}\psi
\end{equation}
is the perturbation term.

The canonical momenta are given then by the well-known formula
\begin{equation}\begin{aligned}
\Pi &= \frac{\partial L}{\partial \partial_0\overline{\psi}}=(\partial^0+isA^0)\psi\\
\overline{\Pi} &= \frac{\partial L}{\partial \partial_0\psi}=(\partial^0-isA^0)\overline{\psi},
\end{aligned}\end{equation}
and the Hamiltonian can be expressed as
\begin{equation}
\mathcal{H}_0=\Pi\partial_0\psi+\overline{\Pi}\partial_0\overline{\psi}-\mathcal{L}_0=\overline{\Pi}\Pi+\overline{(\nabla_\alpha+iA_\alpha)\psi}(\nabla^\alpha+iA^\alpha)\psi.
\end{equation}
In the following discussion $\mathcal{H}_0$ will be taken as the "unperturbed" Hamiltonian density for gravitons. In the next section, we
examine the "unperturbed" equation obtained by taking the variation of the Lagrangian $\mathcal{L}_{KG}$.

\section{The "unperturbed" equation. Separation of variables}
The Euler-Lagrange equations can be obtained by using either (\ref{eq:eulag}) or the well known tensorial - and thus simpler - formula
\[\nabla_i\frac{\partial\mathcal{L}_0}{\partial\nabla_i\overline{\psi}}-\frac{\partial\mathcal{L}_0}{\partial\overline{\psi}}=0.\]
The resulting equation, as mentioned above, is formally a Klein--Gordon equation with an external electromagnetic four-potential:
\begin{equation}
(\nabla_i-isA_i)(\nabla^i-isA_i)\psi=0
\end{equation}
In the following paragraphs, we examine this equation in detail. An explicit computation yields
%
\begin{equation}\label{eq:teudiff}\begin{aligned}
\left(\left((2M-r)r-a^2\right)\psi s^2+\frac{\partial^2\psi}{\partial\varphi^2}-4\frac{\partial^2\psi}{\partial\varphi\partial t}aMr-(a^2-2Mr+r^2)\frac{\partial^2\psi}{\partial r^2}\right.&\\
\left.-(a^4-4a^2Mr-r^4)\frac{\partial^2\psi}{\partial t^2}-\left((2M-r)r-a^2\right)\left(2(M-r)\frac{\partial\psi}{\partial r}-\frac{\partial^2\psi}{\partial\vartheta^2}\right)\right)\cos^2 \vartheta &\\
-\left(\left((2M-r)\frac{\partial^2\psi}{\partial\varphi^2}-4\frac{\partial^2\psi}{\partial\varphi\partial t}aM\right)r-(a^2-2Mr+r^2)^2\frac{\partial^2\psi}{\partial r^2}\right.&\\
\left(a^2\cos^4 \vartheta\frac{\partial^2}{\partial t^2}+ias\cos^3 \vartheta\frac{\partial\psi}{\partial t}+\frac{\partial^2\psi}{\partial\vartheta^2}-2(m-r)\frac{\partial\psi}{\partial r}\right)
                                                         \left((2m-r)r-a^2)\right)&\\
+\left.\left((2m+r)a^2+r^2\right)\frac{\partial\psi}{\partial t^2}-\left(ias\frac{\partial\psi}{\partial t}-\sin \vartheta \frac{\partial\psi}{\partial\vartheta}+is\frac{\partial\psi}{\partial\varphi}\right)
                                                          \left((2m-r)r-a^2\right)\cos \vartheta \right) &=0.
\end{aligned}\end{equation}
It can be shown that the variables in the above equation are separable. Let us look for solutions in the form
\begin{equation}\label{eq:ansatz}
\psi(t,r,\vartheta,\varphi)=e^{-i\omega t}R(r)T(\vartheta)e^{im\varphi}.
\end{equation}

The Euler--Lagrange equations can be normalized so, that the coefficient of $\frac{\partial^2\psi}{\partial\vartheta^2}$ will be 1,
similarly to the original Teukolsky equation (\ref{eq:teuk}).


After substituting the ansatz (\ref{eq:ansatz}) into the above equation,
it can be seen, that none of the terms contains mixed derivatives, and the angle dependence can be factorized in the coefficients of the
radial derivatives, and similarly the $r$-dependence can be factorized in the coefficients of the $\vartheta$-derivatives.


The above properties of the equation make the separation of the variables possible. Straightforward but tedious calculation yields for
the radial equation
\begin{equation}\label{eq:radeq}\begin{aligned}
(a^4 - 4a^2Mr + 2a^2r^2 + 4M^2r^2 - 4Mr^3 + r^4)&R''(r)\\
+2( - a^2M + a^2r + 2M^2r - 3Mr^2 + r^3))&R'(r)=\lambda_r R(r),
\end{aligned}\end{equation}
where
\[\lambda_r=a^4\omega^2 - 4a^2M\omega^2r - a^2m^2 - a^2s^2 - 4aMm\omega r + 2Mrs^2 - \omega^2r^4 - r^2s^2-k\,,\]
in which $k$ is the separation constant.

Similarly the angular equation is
\begin{equation}\label{eq:theq}
(\cos^2 \vartheta  - 1)T''(\vartheta) - \cos \vartheta \sin \vartheta T'(\vartheta)=\lambda_\vartheta T(\vartheta,)
\end{equation}
where
\[\lambda_\vartheta=((\cos \vartheta a\omega-s)\cos^2 \vartheta a\omega+(a\omega + m)s)\cos \vartheta +m^2 + s^2 - a^2\omega^2+k.\]

Introducing $x=\cos \vartheta$ as a new variable in the angular equation we get
\begin{equation}\label{eq:xeq}
(- x^4 + 2x^2 -1)T''(x)+2x(1 - x^2)T'(x)=\lambda_\vartheta T(t).
\end{equation}

Both equations are boundary value problems for second order linear differential operators, with regular boundary conditions
at the horizon and at $r=\infty$.
Unfortunately we have not succeeded to express the solutions of equations (\ref{eq:radeq}) and (\ref{eq:xeq}) using
elementary functions.
Let $\psi_k$ denote the kth eigenfunction of (\ref{eq:teudiff}),
%
then the general solution can be written as
\begin{equation}\label{eq:gensol}
\psi=\sum_k a_k \psi_k.
\end{equation}

Now the canonical quantization of the unperturbed problem is completely straightforward. The modes (\ref{eq:gensol}) of equation (\ref{eq:teudiff})
can be interpreted as gravitons bound to the black hole. 
Defining the corresponding $a_k$ and $a^\dagger_k$ absorption and emission operators for each mode, the Hamiltonian can be
expressed as 
\begin{equation}\label{eq:hamil}
\mathcal{H}=\sum_k E_k a^\dagger_k a_k.
\end{equation}
The $E_k$ energy levels can be determined numerically, as well as the eigenmodes.

Taking into account the complex part of the Hamiltonian (\ref{eq:plag}) of the full Teukolsky equation, the "gravitons" of (\ref{eq:hamil})
become unstable.
This may not come as a complete surprise, as we know that the Kerr black hole admits only quasi normal modes.

The decay constants for these excitations can be calculated using the matrix elements of the perturbation part of the Lagrangian, similarly to
the way decaying states are described with an imaginary energy contribution in nuclear physics.

\section*{Appendix}
In this paper, we quantize the perturbations of a Kerr black hole. We use the Boyer--Lindquist coordinate system: the coordinates are the time,
$t$, the radius,  $r$, and the usual spherical angles, $\vartheta$ and $\varphi$. Using these coordinates the line element can be expressed as
\begin{equation}\label{eq:BLds}\begin{aligned}
ds^2 = & \left(1-\frac{2Mr}{\Sigma}\right)dt^2+\frac{4Mar\sin^2 \vartheta }{\Sigma}dtd\varphi - \frac{\Sigma}{\Delta}dr^2-\Sigma d\vartheta^2\\
 &-\sin^2 \vartheta \frac{r^2+a^2+2Ma^2r\sin^2 \vartheta }{\Sigma}d\varphi^2,
\end{aligned}\end{equation}
where we used
\[\Sigma = r^2+a^2\cos^2 \vartheta \]
and
\[\Delta = r^2-2Mr+a^2\]
for simplicity. Here $M$ denotes the mass of the black hole, and
\[a=\frac{J}{M}\]
is the fraction of its angular momentum ($J$) and mass. It's an important property of these coordinates when deriving the Teukolsky
equation, that these handle the two double principal directions of the space-time symmetrically.

To derive the Teukolsky equation, the Newman--Penrose null tetrad method is utilized.
When using Boyer--Lindquist coordinates, a usual choice for the null tetrad is the Kinnersley tetrad:
\begin{equation}
  \label{eq:kintet}
  \begin{aligned}
(l^i)     &= (\frac{r^2+a^2}{\Delta},1,0,\frac{a}{\Delta})\\
(n^i)     &= \frac{1}{2\Sigma}(r^2+a^2,-\Delta,0,a)\\
(m^i)     &= \frac{1}{\sqrt{2}(r+ia\cos \vartheta )}(ia\sin \vartheta,0,1,\frac{i}{\sin \vartheta}).
  \end{aligned}
\end{equation}
Using these, the non-vanishing spin coefficients are
\begin{equation}
  \label{eq:nabNPBL}
  \begin{aligned}
\rho &= -\frac{1}{r-ia\cos \vartheta }\\
\beta &= \overline{\rho}\frac{\ctg \vartheta }{2\sqrt{2}}\\
\pi &= ia\rho^2\frac{\sin \vartheta }{\sqrt{2}}\\
\tau &= -ia\bar{\rho}\rho\frac{\sin \vartheta }{\sqrt{2}}\\
\mu &= \rho^2\bar{\rho}\frac{\Delta}{2}\\
\gamma &= \mu+\bar{\rho}\rho\frac{r-2M}{2}\\
\alpha &= \pi-\bar{\beta}.
  \end{aligned}
\end{equation}

Here we recall the basic properties of the Teukolsky equation \cite{Teuk1}. This equation is obtained using the Newman--Penrose tetrad
method. This calculation can in fact be done for any space-time in the D Petrov class (ie.\ two pairs of degenerate principal directions).

The perturbations of a Kerr space-time are described by the Teukolsky equation
\begin{equation}
  \label{eq:fteuk}
  \begin{aligned}
\left[\frac{(r^2+a^2)^2}{\Delta}-a^2\sin^2 \vartheta \right]&\frac{\partial^2\psi}{\partial t^2}+\frac{4Mar}{\Delta}\frac{\partial^2\psi}{\partial t\partial\varphi}
                                                               +\left[\frac{a^2}{\Delta}-\frac{1}{\sin^2 \vartheta }\right]\frac{\partial^2\psi}{\partial\varphi^2}\\
-\Delta^{-s}\frac{\partial}{\partial r}\left(\Delta^{s+1}\frac{\partial\psi}{\partial r}\right)&-\frac{1}{\sin \vartheta }\frac{\partial}{\partial\vartheta}\left(\sin \vartheta \frac{\partial
                                                               \psi}{\partial
    \vartheta}\right) -2s\left[\frac{a(r-M)}{\Delta}+\frac{i\cos \vartheta}{\sin^2 \vartheta }\right]\frac{\partial\psi}{\partial\varphi}\\
-2s&\left[\frac{M(r^2-a^2)}{\Delta}-r-ia\cos \vartheta\right]\frac{\partial\psi}{\partial t}+(s^2\ctg^2 \vartheta -s)\psi=4\pi\Sigma T
\end{aligned}
\end{equation}
then. The calculation of $T$ from the energy-momentum tensor of other fields is not included here, but it can be found in Teukolsky's
original paper \cite{Teuk1}.

In this equation, the variables can be separated, using an ansatz
\[\psi=e^{-i\omega t}e^{im\varphi}S(\theta)R(r)\]
for the dependent variable. The radial equation is
\begin{equation}
  \label{eq:teukrad}
  \Delta^{-s}\frac{d}{dr}\left(\Delta^{s+1}\frac{dR}{r}\right)+\left(\frac{K^2-2is(r-M)K}{\Delta}+4is\omega r-\lambda\right)R=0,
\end{equation}
and the angular equation is
\begin{equation}
  \label{eq:teukth}
  \begin{aligned}
  \frac{1}{\sin \vartheta }&\frac{d}{d\vartheta}\left(\sin \vartheta \frac{dS}{d\vartheta}\right)\\
  &+\left(a^2\omega^2\cos^2 \vartheta -\frac{m^2}{\sin^2 \vartheta }-2a\omega s\cos \vartheta-\frac{2ms\cos \vartheta}{\sin^2 \vartheta }-s^2\ctg^2 \vartheta +s+A\right)S=0,
\end{aligned}\end{equation}
then, where
\[K=(r^2+a^2)\omega-am\]
and
\[\lambda=A+a^2\omega^2-2am\omega.\]
The boundary conditions demand $\psi$ being regular at $\vartheta=0,\pi$. The solutions are examined  in a paper by Mano, Suzuki and Takasugi \cite{MSTak}.

According to Bini, Cherubini, Jantzen and Ruffini \cite{Bini}, the above equation is the direct generalization of the Klein--Gordon
equation for arbitrary $s$ spin weight. In their paper, they have brought the equation into a form closely resembling the Klein--Gordon equation
\begin{equation}\label{eq:fBini}
\left[(\nabla^i+s\Gamma^i)(\nabla_i+s\Gamma_i)-4s^2\Psi_2^A\right]\psi=4\pi T,
\end{equation}
where $\Gamma$ is the following four-vector:
\begin{equation}\label{eq:fgamma}\begin{aligned}
\Gamma^t &= -\frac{1}{\Sigma}\left[\frac{M(r^2-a^2)}{\Delta}-(r+ia\cos \vartheta)\right]\\
\Gamma^r &= -\frac{1}{\Sigma}(r-M)\\
\Gamma^\vartheta &= 0\\
\Gamma^\varphi &= -\frac{1}{\Sigma}\left[\frac{a(r-M)}{\Delta}+i\frac{\cos \vartheta}{\sin^2 \vartheta }\right].
\end{aligned}\end{equation}

\section*{Acknowledgements}
\'A.~L.\ would like to thank his supervisor, Zolt\'an~Perj\'es for all what
he has learnt from him, and for suggesting this problem. The tragic loss of Zolt\'an is a great blow for all Hungarian physics.

\'A.~L.\ also would like to thank Gy\"orgy P\'ocsik and P\'eter Gn\"adig for discussions, and P\'eter Forg\'acs for his help in preparing
this manuscript.

This work has been supported by OTKA grant no.\ TS044665.


\begin{thebibliography}{17}

\bibitem{Gupta1}
S.~N. Gupta, \emph{Proc. Phys. Soc. A}, \textbf{65}, 161--169
  (1952).\\
S.~N. Gupta, \emph{Proc. Phys. Soc. A}, \textbf{65}, 608--619
  (1952).


\bibitem{ADM}
R.~L. Arnowitt, S.~Deser, and C.~W. Misner, "{Canonical Analysis of
  General Relativity}", in \emph{{Recent Developments in General Relativity}},
  1962.\\
R.~L. Arnowitt, S.~Deser, and C.~W. Misner, "{The Dynamics of General
  Relativity}", in \emph{{Gravitation: An Introduction to Current Research}},
  edited by L.~Witten, 1962.

\bibitem{Penrose}
R.~Penrose, \emph{{Phys. Reports}}, \textbf{6}, 241--316 (1972).\\
R.~Penrose, "{Twistor theory: it's aims and achievements}", in \emph{Quantum
   Gravity: An Oxford Symposium }, eds. C.~J.~Isham, R.~Penrose, D.~W.~Sciama, Calrendon Press, Oxford, 1975. 

\bibitem{Ashtekar}
A.~Ashtekar, \emph{Phys. Rev.}, \textbf{D36}, 1587--1602 (1987).\\
A.~Ashtekar, \emph{{Lectures on non-perturbative canonical gravity}}, World
  Scientific, 1991.

\bibitem{Rovelli}
C.~Rovelli, \emph{{Loop Quantum Gravity}}, Living Rev. Relativity, \textbf{1}, 1998.


\bibitem{Teuk1}
S.~A. Teukolsky, \emph{ApJ}, \textbf{185}, 635--647 (1973).\\
S.~A. Teukolsky, and W.~H. Press, \emph{ApJ}, \textbf{185}, 649--672 (1973).

\bibitem{Chrz}
P.~L. Chrzanowski, \emph{Phys. Rev.}, \textbf{D11}, 2042--2062 (1975).\\
P.~L. Chrzanowski, and C.~W. Misner, \emph{Phys. Rev.}, \textbf{D10},
  1701--1721 (1974).

\bibitem{Ori}
A.~Ori, \emph{Phys. Rev.}, \textbf{D67}, 124010 (2003).

\bibitem{Anco1}
S.~Anco, {Conservation Laws of the Teukolsky equation for massless spin $s$
  fields in {Kerr} spacetime} (2003).

\bibitem{Bini}
D.~Bini, C.~Cherubini, R.~Jantzen, and R.~Ruffini, \emph{Prog.Theor.Phys.},
  \textbf{107}, 967--992 (2002).


%
\bibitem{MSTak}
S.~Mano, H.~Suzuki, and E.~Takasugi, \emph{Prog. Theor. Phys.}, \textbf{95},
  1079--1096 (1996).

\end{thebibliography}
\end{document}